\documentclass[notitlepage,pallis,twoside]{wpallis}
\usepackage{fancyh}
\usepackage[l]{floatflt}
\usepackage{epsfig}
\usepackage{amssymb}
\usepackage{latexsym}
\usepackage{times}
\usepackage{slashed,cite}
\numberwithin{equation}{section}

\newcommand{\eem}{\end{matrix}}
\newcommand{\bem}{\begin{matrix}}
\newcommand{\eeq}{\end{equation}}
\newcommand{\beq}{\begin{equation}}
\newcommand{\ba}{\begin{array}}
\newcommand{\ea}{\end{array}}
\newcommand{\bea}{\begin{eqnarray}}
\newcommand{\eea}{\end{eqnarray}}
\newcommand{\baq}{\begin{eqnarray}}
\newcommand{\eaq}{\end{eqnarray}}

\newcommand\eq[1]{Eq.~(\ref{#1})}
\newcommand\eqs[2]{Eqs.~(\ref{#1}) and (\ref{#2})}
\newcommand\eqss[3]{Eqs.~(\ref{#1}), (\ref{#2}) and (\ref{#3})}

\newcommand{\ftn}{\footnotesize}

\newcommand{\GeV}{{\mbox{\rm GeV}}}
\newcommand{\sFig}[2]{Fig.~\ref{#1}-${\sf ({#2})}$}

\newcommand{\etal}{{\it et al.\/}}

\newcommand{\Vhi}{\ensuremath{\hat V_{0}}}
\newcommand{\ck}{\ensuremath{c_{4K}}}
\newcommand{\kf}{\ensuremath{k_{\rm 4}}}
\newcommand{\ks}{\ensuremath{k_{\rm 6}}}
\newcommand{\ns}{\ensuremath{n_{\rm s}}}
\newcommand{\Dex}{\ensuremath{\Delta_{\rm m*}}}
\newcommand{\ai}{\ensuremath{\alpha_m}}
\newcommand{\bi}{\ensuremath{\beta_m}}
\newcommand{\hm}{\ensuremath{h_m}}

\newcommand{\phh}{\ensuremath{\Phi}}
\newcommand{\bph}{\ensuremath{\bar \Phi}}

\def\M{{\bar{M}}}
\def\N{{\bar{N}}}

\def\W{{\hat{W}}}
\def\K{{\hat{K}}}
\def\Kt{{\tilde{K}}}

\def\Z{\hat{Z}}
\def\n{\bar{n}}

\def\Ms{\hat{M}_{\rm S}}

\def\ä{\"{a}}

\def\p{|S|}

\begin{document}

\thispagestyle{empty}

\title[]{\Large\bfseries\scshape K\"ahler Potentials for Hilltop F-term Hybrid Inflation}

\author{\large\bfseries\scshape C. Pallis}
\address[]{\sl Department of Physics, University of Patras,
\\ GR-265 00 Patras,  GREECE \\{\tt kpallis@auth.gr}}

\begin{abstract}{\small {\bfseries\scshape Abstract} \\
\par We consider the basic supersymmetric (SUSY)
models of F-term hybrid inflation (FHI). We show that a simple
class of K\"ahler potentials ensures a resolution to the $\eta$
problem and allows for inflation of hilltop type. As a
consequence, observationally acceptable values for the spectral
index, $\ns$, can be achieved constraining the coefficient $\ck$
of the quartic supergravity correction to the inflationary
potential. For about the central value of $\ns$, in the case of
standard FHI, the grand unification (GUT) scale turns out to be
well below its SUSY value with the relevant coupling constant
$\kappa$ in the range $(0.0006-0.15)$ and $\ck\simeq-(1100-0.05)$.
In the case of shifted [smooth] FHI, the GUT scale can be
identified with its SUSY value for $\ck\simeq-16$
[$\ck\simeq-1/16$].}
%
\end{abstract} \maketitle
\publishedin{{\sl J. Cosmol. Astropart. Phys.} {\bf 04}, 024
(2009)}

\thispagestyle{empty}

\setcounter{page}{1} \pagestyle{fancyplain}


\rhead[\fancyplain{}{ \bf \thepage}]{\fancyplain{}{\sl K\"ahler
Potentials for Hilltop F-term Hybrid Inflation}}
\lhead[\fancyplain{}{\sl \leftmark}]{\fancyplain{}{\bf \thepage}}
\cfoot{}

\section{Introduction}\label{intro}

One of the most natural and well-motivated classes of inflationary
models is the class of \emph{supersymmetric} (SUSY) \emph{F-term
hybrid inflation} (FHI) models \cite{hybrid}. In particular, the
basic versions of FHI are the standard \cite{susyhybrid}, shifted
\cite{jean} and smooth \cite{pana1} FHI. They are
realized~\cite{susyhybrid} at (or close to) the SUSY \emph{Grand
Unified Theory} (GUT) scale $M_{\rm
GUT}\simeq2.86\cdot10^{16}~{\rm GeV}$ and can be easily linked to
several extensions \cite{lectures} of the \emph{Minimal
Supersymmetric Standard Model} (MSSM) which have a rich structure.
Namely, the $\mu$-problem of MSSM is solved via a direct coupling
of the inflaton to Higgs superfields \cite{dvali} or via a
Peccei-Quinn symmetry \cite{rsym}, baryon number conservation is
an automatic consequence \cite{dvali} of an R symmetry and the
baryon asymmetry of the universe is generated via leptogenesis
which takes place \cite{lept} through the out-of-equilibrium
decays of the inflaton's decay products.

Although quite successful, these models have at least two
shortcomings: (i) the so-called $\eta$ problem and (ii) the
problem of the enhanced (scalar) spectral index, $n_{\rm s}$. The
first problem is tied \cite{hybrid, review, eta} on the
expectation that \emph{supergravity} (SUGRA) corrections generate
a mass squared for the inflaton of the order of the Hubble
parameter during FHI and so, the $\eta$ criterion is generically
violated, ruining thereby FHI. Inclusion of SUGRA corrections with
canonical K\"ahler potential prevents \cite{hybrid, senoguz} the
generation of such a mass term due to a mutual cancellation.
However, despite its simplicity, the canonical K\"ahler potential
can be regarded \cite{hybrid} as fine tuning to some extent and
increases, in all cases, even more $n_{\rm s}$. This aggravates
the second problem of FHI, i.e., the fact that, under the
assumption that the problems of the \emph{standard big bag
cosmology} (SBB) are resolved exclusively by FHI, these models
predict $n_{\rm s}$ just marginally consistent with the fitting of
the five-year results \cite{wmap} from the \emph{Wilkinson
Microwave Anisotropy Probe Satellite} (WMAP5) data with the
standard power-law cosmological model with cold dark matter and a
cosmological constant ($\Lambda$CDM). According to this, $n_{\rm
s}$ at the pivot scale $k_*=0.002/{\rm Mpc}$ is to satisfy
\cite{wmap} the following range of values:
\begin{equation}\label{nswmap}
n_{\rm s}=0.963^{+0.014}_{-0.015}~\Rightarrow~0.933\lesssim n_{\rm
s} \lesssim 0.991\>\>\mbox{at 95$\%$ confidence level}.
\end{equation}

\tableofcontents\vskip-1.3cm\noindent\rule\textwidth{.4pt}\\

One possible resolution (for other proposals, see \cref{battye,
mhi}) of the tension between FHI and the data is \cite{gpp, king}
the utilization of a quasi-canonical \cite{CP} K\"ahler potential
with a convenient choice of the sign of the next-to-minimal term.
As a consequence, a negative mass term for the inflaton is
generated. In the largest part of the parameter space the
inflationary potential acquires a local maximum and minimum. Then,
FHI of the hilltop type \cite{lofti} can occur as the inflaton
rolls from this maximum down to smaller values. This set-up
provides acceptable values for both $\eta$ and $\ns$ but it
requires \cite{king, gpp, hinova} two kinds of mild tuning: (i)
the relevant coefficient in the K\"ahler potential is to be
sufficiently low (ii) the value of the inflaton field at the
maximum is to be sufficiently close to the value that this field
acquires when the pivot scale crosses outside the inflationary
horizon.

In this paper, we propose a class of K\"ahler potentials which
supports a new type of hilltop FHI (driven largely by the quartic
rather than the quadratic SUGRA correction) without the first kind
of tuning above. In particular, the coefficients of K\"ahler
potentials are constrained to natural values (of order unity) so
as the mass term of the inflaton field is identically zero. The
achievement of the observationally acceptable $\ns$'s requires a
mild tuning of the initial conditions similar to that needed in
the case with quasi-canonical K\"ahler potential. The suggested
here form of K\"ahler potentials has been previously proposed in
\cref{nurmi} in order to justify the saddle point condition needed
for the attainment of $A$-term or MSSM inflation \cite{Aterm}. A
similar idea is also explored in \cref{sugraP} without, though,
the $\ns$ problem to be taken into account.

Below, we describe the proposed embedding of the basic FHI models
in SUGRA (Sec.~\ref{sugra}) and we derive the inflationary
potential (Sec.~\ref{sec:Vhi}). Then we exhibit the observational
constraints imposed on our models (Sec.~\ref{obs}) and we end up
with our numerical results (Sec.~\ref{res}) and our conclusions
(Sec.~\ref{con}). Throughout the text, we set $\hbar=c=k_{\rm
B}=1$. Hereafter, parameters with mass dimensions are measured in
units of the reduced Planck mass ($m_{\rm
P}=2.44\cdot10^{18}~\GeV$) which is taken to be unity.

\section{FHI in non-Minimal SUGRA}\label{sugra}

In this section we outline the salient features of our set-up
(Sec.~\ref{sugra1}), we extract the SUSY potential
(Sec.~\ref{sugra2}), we calculate the SUGRA corrections
(Sec.~\ref{sugra3}) and present the proposed class of K\"ahler
potentials (Sec.~\ref{sugra4}).

\subsection{The General Set-up}\label{sugra1}

The F-term hybrid inflation can be realized within SUGRA adopting
one of the superpotentials below:
\begin{equation} \label{Whi} W = \W + W_{\rm FHI}\>\>\mbox{with}\>\>W_{\rm FHI}=\left\{\begin{matrix}
\hat\kappa S\left(\bar \Phi\Phi-M^2\right)\hfill   & \mbox{for
standard FHI}, \hfill \cr
\hat\kappa S\left(\bar \Phi\Phi-M^2\right)-S{(\bar
\Phi\Phi)^2\over\Ms^2}\hfill  &\mbox{for shifted FHI}, \hfill \cr
S\left({(\bar \Phi\Phi)^2\over \Ms^2}-\hat\mu_{\rm
S}^2\right)\hfill &\mbox{for smooth FHI}. \hfill \cr\end{matrix}
\right. \end{equation}
Here we use the hat to denote quantities (such as the part $\W$ of
$W$) which depend exclusively on the hidden sector superfields,
$\hm$. Also, $\bar{\Phi}$ and $\Phi$ is a pair of left handed
superfields belonging to non-trivial conjugate representations of
a GUT gauge group $G$ and reducing its rank by their \emph{vacuum
expectation values} (v.e.vs), $S$ is a gauge singlet left handed
superfield, $\Ms\sim 0.205$ is an effective cutoff scale
comparable with the string scale and the parameters $\hat\kappa$
and $M,~\hat\mu_{\rm S}~(\sim M_{\rm GUT}=4.11\cdot10^{-3})$ are
made positive by field redefinitions.

$W_{\rm FHI}$ in Eq.~(\ref{Whi}) for standard FHI is the most
general renormalizable superpotential consistent with a continuous
R symmetry \cite{susyhybrid} under which
\begin{equation}
  \label{Rsym}
S\  \rightarrow\ e^{ir}\,S,~\Phi\ \rightarrow\
e^{ir}\,\Phi,~\bar\Phi\ \rightarrow\ e^{-ir}\Phi,~W \rightarrow\
e^{ir}\, W.
\end{equation}
Including in this superpotential the leading non-renormalizable
term, one obtains $W_{\rm FHI}$ of shifted \cite{jean} FHI in
Eq.~(\ref{Whi}). Finally, $W_{\rm FHI}$ of smooth \cite{pana1} FHI
can be produced if we impose an extra $Z_2$ symmetry under which
$\Phi\rightarrow -\Phi$ and, therefore, only even powers of the
combination $\bar{\Phi}\Phi$ can be allowed.

To keep our analysis as general as possible, we do not adopt any
particular form for $\W$ (for some proposals see \cref{martin,
covi,davis}). Note that our construction remains intact even if we
set $\W=0$ as it was supposed in \cref{sugraP}. This is due to the
fact that $\W$ is expected to be much smaller than the
inflationary energy density (see \Sref{sugra3}). For $\W\neq0$,
though, we need to assume \cite{nurmi} that $h_m$'s are stabilized
before the onset of FHI by some mechanism not consistently taken
into account here \cite{lalak}. As a consequence, we neglect the
dependence of $\W$, $\hat\kappa$, $\hat\mu_{\rm S}$ and $\Ms$ on
$h_m$ and so, these quantities are treated \cite{nurmi} as
constants. We further assume that the D-terms due to $h_m$'s
vanish (contrary to the strategy followed in \cref{sugraP}).

The SUGRA scalar potential (without the D-terms) is given (see,
e.g., \cref{review, martin}) by
\begin{equation}
V_{\rm SUGRA}=e^{K}\left(K^{M\N}F_M\, F^*_\N -3\vert
W\vert^2\right)\>\>\mbox{where}\>\>F_M=W_M +K_M W \label{Vsugra}
\end{equation}
is the SUGRA generalization of the F-terms, the subscript $M~[\M]$
(not to be confused with the parameter $M$ in \eq{Whi}) denotes
derivation \emph{with respect to} (w.r.t) the complex scalar field
$\phi_M~[\phi_M^{*}]$ which corresponds to the chiral superfield
$\phi_M$ with $\phi_M=h_m,S,\phh,\bph$ and the matrix $K^{M\N}$ is
the inverse of the K\"ahler metric $K_{M\N}$. In this paper we
consider a quite generic form of K\"{a}hler potentials, which
respect the R symmetry of \Eref{Rsym}. Namely we take
\beq \label{K}
K=\K+\Z\p^2+{1\over4}\kf\Z^2\p^4+{1\over6}\ks\Z^3\p^6+|\phh|^2+|\bph|^2,
\eeq
where $\kf$ and $\ks$ are positive or negative constants of order
unity and the functions $\K$ and $\Z$ are to be determined. The
non-vanishing entries of $K^{M\N}$ are
\numparts \baq K^{m\n}&
\simeq&\K^{m\n}-\Kt^{m\n}\p^2\>\>\mbox{with}\>\>\Kt^{m\n}=\K^{m{\tilde\n}}\K^{\tilde
m{\n}}\left(Z_{\tilde m\tilde\n}-\Z_{\tilde
m}\Z_{\tilde\n}/\Z\right),\\  K^{mS^*} \hspace*{-0.1cm}&\simeq&
\left(\Kt^{m\n}\Z_{\n}S^*\p^2-\Z^m S^*\right)/\Z, \\
K^{S\n}\hspace*{0.1cm}  &\simeq&\left(\Kt^{m\n}\Z_m S\p^2-\Z^{\n} S\right)/\Z,\\
 K^{SS^*}&\simeq&1/\Z+\left(\Z^m\Z_m/\Z^2-\kf\right)\p^2+
\left[\left(\kf^2-3\ks/2\right)\Z-\Kt^{m\n}\Z_m\Z_{\n}/\Z^2\right]\p^4,\\
 K^{\phh\phh^*}&=&1\>\>\mbox{and}\>\>K^{\bph\bph^*}=1,
\eaq\endnumparts \hspace{-.14cm}
where the indices $m$ and $n$ are raised and lowered with
$\K^{m\bar n}$ and we keep only the terms necessary in order to
extract a reliable expansion of $V_{\rm SUGRA}$ up to the order
$\p^4$ (see \Sref{sugra3}).

\subsection{The SUSY Potential}\label{sugra2}

The SUSY potential includes \cite{review, martin} F- and D-term
contributions. Note that, as a consequence of our assumptions
about the nature of $\bar{\Phi}$ and $\Phi$ and the structure of
$K$ in \eq{K}, the D-term contribution vanishes for
$\vert\bar{\Phi} \vert=\vert\Phi\vert$. Expanding $V_{\rm SUGRA}$
in \Eref{V} for $\p\ll1$ and $W\ll1$ and neglecting soft SUSY
breaking terms (see, e.g., \cref{martin}), we can extract the
F-term contribution to the SUSY potential, which can be written as
\beq \label{VF} V_{\rm F}\simeq\left\{\bem
\kappa^2M^4\left(({\sf \Phi}^2-1)^2+2{\sf S}^2{\sf
\Phi}^2\right)\hfill & \mbox{for standard FHI}, \hfill \cr
\kappa^2M^4\left(({\sf \Phi}^2-1-\xi{\sf \Phi}^4)^2+2{\sf S}^2{\sf
\Phi}^2(1-2\xi{\sf \Phi}^2)^2\right)\hfill &\mbox{for shifted
FHI}, \hfill \cr
\mu^4_{\rm S}\left(({\sf \Phi}^4-1)^2+8{\sf S}^2{\sf
\Phi}^6\right) \hfill &\mbox{for smooth FHI}, \hfill \cr\eem
\right.\eeq
where $\xi=M^2/\kappa M^2_{\rm S}$ with \cite{jean}
$1/7.2<\xi<1/4$. In order to recover the properly normalized
energy density during FHI (see below), we absorb in \eq{VF} some
normalization pre-factors emerging from $V_{\rm SUGRA}$, defining
the quantities $\kappa=e^{\K/2}\Z^{-1/2}\hat\kappa$ and $\mu_{\rm
S}=e^{\K/4}\Z^{-1/4}\hat\mu_{\rm S}$. We then define $M_{\rm
S}=e^{-\K/4}\Z^{1/4}\Ms$ so as $\kappa M^2_{\rm S}=\hat\kappa \hat
M^2_{\rm S}$ and $\mu_{\rm S} M_{\rm S}=\hat\mu_{\rm S}\Ms$. Also,
we use \cite{jean, pana1} the dimensionless quantities:
\beq\left\{\bem
{\sf \Phi}=|\Phi|/M~~\mbox{and}~~{\sf S}=\Z^{1/2}|S|/M\hfill &
\mbox{for standard or shifted FHI,} \hfill\cr
{\sf \Phi}=|\Phi|/\sqrt{\mu_{\rm S} M_{\rm S}}~~\mbox{and}~~{\sf
S}=\Z^{1/2}|S|/\sqrt{\mu_{\rm S} M_{\rm S}}\hfill &\mbox{for
smooth FHI.} \hfill \cr\eem
\right.\eeq
Recall that the scalar components of the superfields are denoted
by the same symbols as the corresponding superfields.

The potential in \Eref{VF} reveals that $W_{\rm FHI}$ in
Eq.~(\ref{Whi}) plays a twofold crucial role:

\begin{itemize}

\item It leads to the spontaneous breaking of $G$. Indeed, the
vanishing of $V_{\rm F}$ gives the v.e.vs of the fields in the
SUSY vacuum. Namely,
\begin{equation} \label{vevs} \langle S\rangle=0~~\mbox{and}~~\vert\langle\bar{\Phi}
\rangle\vert=\vert\langle\Phi\rangle\vert=v_{_G}=\left\{\bem
M\hfill   & \mbox{for standard FHI}, \hfill \cr
\frac{M\sqrt{1-\sqrt{1-4\xi}}}{\sqrt{2\xi}}\hfill  &\mbox{for
shifted FHI}, \hfill \cr
\sqrt{\mu_{\rm S}M_{\rm S}}\hfill  &\mbox{for smooth FHI} \hfill
\cr\eem
\right. \end{equation}
(in the case where $\bar{\Phi}$, $\Phi$ are not {\it Standard
Model} (SM) singlets, $\langle\bar{\Phi} \rangle$, $\langle{\Phi}
\rangle$ stand for the v.e.vs of their SM singlet directions).

\item It gives rise to FHI. This is due to the fact that, for
large enough values of $|S|$, there exist valleys of local minima
of the classical potential with constant (or almost constant in
the case of smooth FHI) values of $V_{\rm F}$. In particular, we
can observe that $V_{\rm F}$ takes the following constant value
\beq V_{\rm HI0}=\left\{\bem
\kappa^2 M^4\hfill \cr
\kappa^2 M_\xi^4\hfill \cr
\mu_{\rm S}^4\hfill \cr \eem
\right.\>\>\mbox{along the direction(s):}\>\>{\sf \Phi}=\left\{
\bem
0~~\hfill & \mbox{for standard FHI}, \hfill\cr
0\>\>\mbox{or}\>\>1/\sqrt{2\xi} \hfill &\mbox{for shifted FHI},
\hfill \cr
0\>\>\mbox{or}\>\>1/\sqrt{6}{\sf S} \hfill &\mbox{for smooth FHI},
\hfill \cr\eem
\right.\eeq
with $M_\xi=M\sqrt{1/4\xi-1}$. It can be shown \cite{hinova} that
the flat direction ${\sf \Phi}=0$ corresponds to a minimum of
$V_{\rm F}$, for $|S|\gg M$, in the cases of standard and shifted
FHI, and to a maximum of $V_{\rm F}$ in the case of smooth FHI. As
a consequence, topological defects such as strings \cite{jp,
mairi, gpp}, monopoles, or domain walls may be produced
\cite{pana1} via the Kibble mechanism \cite{kibble} during the
spontaneous breaking of $G$ at the end of standard FHI, since this
type of FHI can be realized only for ${\sf \Phi}=0$. On the
contrary, this can be avoided in the other two cases, since the
form of $W_{\rm FHI}$ allows for non-trivial inflationary valleys
of minima with ${\sf \Phi}\neq0$, along which $G$ is spontaneously
broken.

\end{itemize}

\subsection{SUGRA Corrections}\label{sugra3}

The consequences that SUGRA has on the models of FHI can be
investigated by restricting ourselves to the inflationary
trajectory $\phh=\bph\simeq0$ (possible corrections due to the
non-vanishing $\phh$ and $\bph$ in the cases of shifted and smooth
FHI are \cite{sstad} negligible). Therefore, $W$ in \Eref{Whi}
takes the form
\beq\label{W} W = \W + I\
,\>\>\mbox{where}\>\>I=-\Vhi^{1/2}S\>\>\mbox{with}\>\>\Vhi=e^{-\K}\Z\,V_{\rm
HI0}. \eeq
Given the superpotential above, the scalar potential in
Eq.~(\ref{Vsugra}) can be written as
\beq \label{V} V_{\rm SUGRA} = |\W|^2V_{\W}+ \W I^*V_{\W I} +
\W^*I V_{\W I}^* + \Vhi V_{I}\ ,~~\mbox{where}\eeq
\vspace{-0.5cm}\numparts\baq \label{Vw} V_{\W} & = & e^K
\left(K^{M\N}K_MK_\N - 3\right)\ ,  \\ \label{Vwi} V_{\W I} & = &
e^K \left(K^{M\N}K_MK_\N + K^{MS^*}K_M/ S^*- 3\right)\ ,\\
\label{Vi} V_{I} & = & e^K \Big[K^{SS^*} + S K^{MS^*}K_M +
S^*K^{S\N} K_\N + |S|^2\left(K^{M\N}K_MK_\N-3\right)\Big]\ .
\eaq\endnumparts \hspace{-.14cm}
Using the K\ähler potential in Eq.~(\ref{K}) we can obtain an
expansion of $V_{\rm SUGRA}$ in powers of $\p$. To this end, we
first expand in powers of $\p$ the involved in Eqs.~(\ref{Vw}) --
(\ref{Vi}) expressions:
\numparts\baq \label{kmn} K^{M\N}K_MK_\N
&\simeq&\K^m\K_m+\p^2\left(\Z-\Kt^{m\n}\K_m\K_{\n}\right)+{\cal
O}(\p^4),\\ \label{ksm}K^{M
S^*}K_M&\simeq&(1-\Z^m\K_m)S^*+S^*\p^2\left(\Kt^{m\n}\K_m\Z_{\n}-\kf
\Z/2\right)/\Z,\\ \label{ksn}K^{S\N}
K_\N&\simeq&(1-\K^m\Z_m)S+S\p^2\left(\Kt^{m\n}\Z_m\K_{\n}-\kf
\Z/2\right)/\Z. \eaq\endnumparts \hspace{-.14cm}
Substituting Eqs.~(\ref{kmn}) -- (\ref{ksn}) into \Eref{V} and
taking into account that
\beq \label{ex} e^K\simeq e^\K\,\left(1+\Z\p^2 +\
(1+\kf/2)\Z^2\p^4/2\right),\eeq
we end up with the following expansion:
\beq \label{Vfinal} V_{\rm SUGRA}\simeq
V_0+V_1\p+V_2\p^2+V_4\p^4\,, ~~\mbox{where}\eeq
\vspace{-0.5cm}\numparts\baq \label{V0} V_{0}&\simeq&e^{\K}\
\Z^{-1}\ \Vhi\ ,\\ \label{V1} V_{1}&\simeq&2e^{\K}\ \Vhi^{1/2}\
|\W| \left(\K^m\K_m-\Z^m\K_m/\Z-2\right)\cos\theta,\\
\label{V2} V_{2}&\simeq&e^{\K}\ \Vhi
\Big[\K^m\K_m-\left(\K^m\Z_m+\Z^m\K_m\right)/\Z+\Z^m\Z_m/\Z^2 -\kf
\Big],\\
\nonumber V_{4}&\simeq&e^{\K}\ \Vhi\
Z^{-1}\left[\Kt^{m\n}\left(\Z_m\K_{\n}+\K_m\Z_{\n}-\Z_m\Z_{\n}/\Z-\Z\K_m\K_{\n}\right)+\Z^m\Z_m\right.\\
&-& \label{V4}\left.\Z\left(\Z^m\K_m+\K^m\Z_m\right)
+\left(\K^m\K_m+{1\over2}-{7\over4}\kf+\kf^2-{3\over2}\ks
\right)\Z^2\right],\eaq\endnumparts \hspace{-.14cm}
where the phase $\theta$ in $V_1$ reads $\theta={\rm
arg}\left(\K^m\K_m-\Z^{m}\K_{m}/\Z-2\right)+{\rm arg}(\W)-{\rm
arg}(\Vhi^{1/2})-{\rm arg}(S)$. In the \emph{right hand side}
(r.h.s) of \eqss{V0}{V2}{V4} we neglect terms proportional to
$|\W|^2$ which are certainly subdominant compared with those which
are proportional to $\Vhi$. From the terms proportional to
$|\W|\Vhi^{1/2}$ we present, just for completeness, the term $V_1$
which expresses the most important contribution \cite{covi, sstad}
to the inflationary potential from the soft SUSY breaking terms.
For natural values of $\W$ and $e^\K$ this term starts
\cite{sstad,jp} playing an important role in the case of standard
FHI for $\kappa\lesssim5\cdot10^{-4}$ whereas it has \cite{sstad}
no significant effect in the cases of shifted and smooth FHI. For
simplicity, we neglect it, in the following. Note, finally, that
the well-known results in the context of minimal \cite{senoguz}
[quasi-minimal \cite{CP, king, hinova}] SUGRA can be recovered
from \eqs{V2}{V4} by setting $\K=0,~\Z=1$ and $\kf=\ks=0$ [$\K=0$
and $\Z=1$].

\subsection{Imposed Conditions}\label{sugra4}

From \eq{Vfinal} we infer that a resolution to the $\eta$ problem
of FHI requires $V_2=0$ - needless to say that there is no
contribution to $\eta$ from the neglected  $V_1$-term in \eq{V1}.
Considering a well motivated, by several superstring and D-brane
models \cite{Ibanez}, form for $\K$ and $\Z$, we can impose
constraints on their parameters and on $\kf$ and $\ks$ so as the
requirement above is fulfilled identically. In particular,
inspired from \cref{sugraP, nurmi}, we seek the following ansatz
for $\K$ and $\Z$  :
\beq \label{k2} \K=\sum_{m=1}^{\sf M}\beta_m
\ln(h_m+h_{m}^{*})\>\>\mbox{and}\>\>\Z=k_Z\prod_{m=1}^{\sf
M}(h_m+h_{m}^{*})^{\alpha_m},\>\>\mbox{with}\>\>\beta=\sum_{m=1}^{\sf
M}\bi<0.\eeq
The latter restriction is demanded so as the exponential of
$V_{\rm SUGRA}$ in \eq{Vsugra} is well defined for $h_m\sim1$. We
further assume that $\beta_m$'s have to be integers and
$\alpha_m$'s have to be rational numbers. Although negative
integers as $\beta_m$'s are more frequently encountered, positive
$\beta_m$'s are also allowed \cite{Lust}. Since ${\sf M}$ measures
the number of hidden sector fields, we restrict ourselves to its
lowest possible values. Inserting \eq{k2} into \eqss{V0}{V2}{V4},
\eq{Vfinal} takes the form
\beq \label{Vfinals} V_{\rm SUGRA}\simeq V_{\rm
HI0}\left(1-{1\over2}c_{2K}\sigma^2+{1\over4}c_{4K}\sigma^4\right),
\>\>\mbox{where}\>\>\sigma=\sqrt{2}\Z^{1/2}\p\eeq
is the canonically (up to the order $\p^2$) normalized inflaton
field and the coefficients $c_{2K}$ and $c_{4K}$ read
\numparts\baq \label{c2k} c_{2K}&=&-{V_2\over
e^{\K}\Vhi}=\kf+\sum_{m=1}^{\sf M}{(\ai-\bi)^2\over\bi}\\
\label{c4k} c_{4K}&=&{V_4\over e^{\K}\Vhi
Z}=\kf^2-{7\over4}\kf-{3\over2}\ks+{1\over2}+\sum_{m=1}^{\sf
M}{(\ai-\bi)^3\over\bi^2}\cdot \eaq\endnumparts
\begin{floatingfigure}[lt]{2.5in}
\begin{center}
\begin{tabular}{|lll||l|}\hline
$-\beta_1$&$-\alpha_1$&$\kf$&$\ck$\\\hline\hline
$1$&$3/2$&$1/4$&$0$\\
$6$&$4$&$2/3$&$0$\\\hline%
$4$&$3$&$1/4$&$3/16$\\
$1$ &$-1/2$&$9/4$&$5$\\ \hline
\end{tabular} \\
\end{center}
\hspace*{0.8cm} {\small{\sl{\bfseries Table 1:}} \sl Solutions to
\eq{ab} for\\ \hspace*{0.8cm} ${\sf M}=1$  and $\ks=0$.}\label{table1}\\
\end{floatingfigure}
Consequently, FHI can be deliberated from the $\eta$ pro-blem if
the following condition is valid:
\beq \label{ab} \hspace*{2.8in} c_{2K}=0.\eeq
On the other hand, the data on $\ns$ favors hilltop FHI which can
be attained \cite{lofti} for $\ck<0$. However, $\ck>0$ is still
marginally allowed. In Table~1 we list solutions to Eq.~(\ref{ab})
for the simplest case with ${\sf M}=1$ and $\ks=0$ with
$\ck\geq0$. Solutions to Eq.~(\ref{ab}) with the observationally
favored $\ck<0$ can be also achieved with a variety of ways. Note,
initially, that $\ks>0$ is beneficial for this purpose, since it
decreases $\ck$, without disturbing the satisfaction of \eq{ab}. A
first set of solutions can be taken for $\ai=0$. In this case
(which resembles the cases studied in \cref{sugraP}) setting,
e.g., $\kf=-\bi=1$, we get $\ck=3/4,~0,~-3,~-6,-9$ for
$\ks=0,1/2,5/2,9/2,13/2$.

\begin{figure}[!t]\vspace*{-.46in}\begin{tabular}[!h]{cc}\begin{minipage}[t]{8in}
\hspace{1.15in}
\epsfig{file=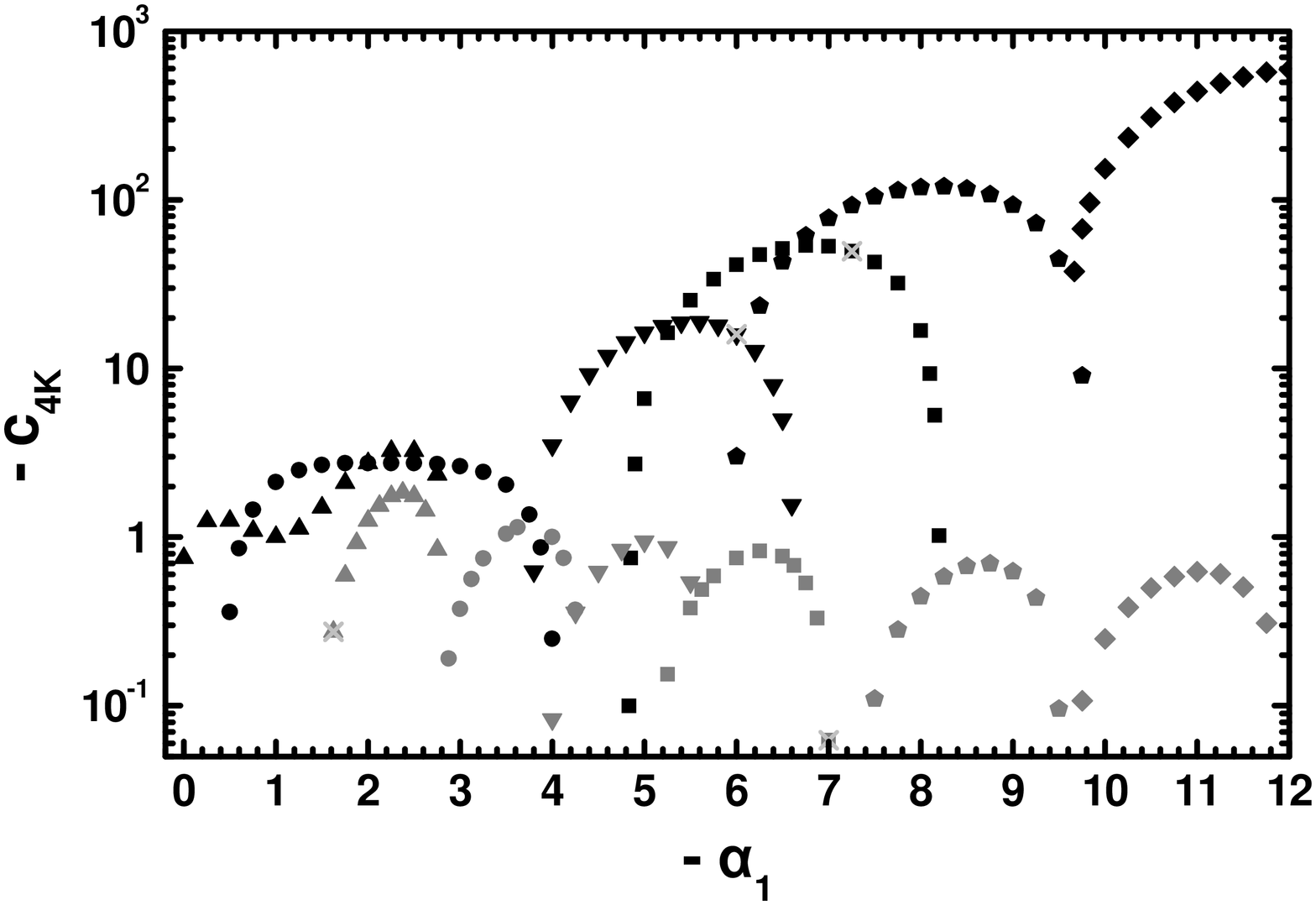,height=3.65in,angle=-90}\end{minipage}
&\begin{minipage}[h]{2in}
\hspace{-3.5in}{\vspace*{-1.8in}\includegraphics[height=3.5cm,angle=-90]
{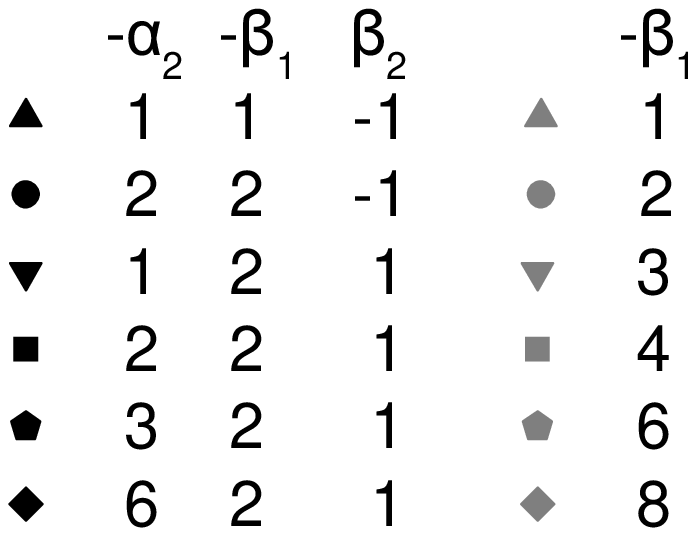}}\end{minipage}
\end{tabular} \hfill \vchcaption[]{\sl \small Values of $c_{4K}$ obtained from \eq{c4k}
versus $\alpha_1$ for ${\sf M}=1$ and $\ks=0$ (gray points) or
${\sf M}=2$ and $\ks=1$ (black points) with
$0.1\lesssim|\kf|\lesssim10$ given by \eqs{c2k}{ab}. Light gray
crosses correspond to $\ck$'s used in Table~2. The adopted values
for the residual parameters ($\ai$ and $\bi$) are also
shown.}\label{fig1}
\end{figure}

More generically, taking as input parameters $\ai$'s and $\bi$'s
we can assure the fulfillment of \eq{ab} constraining $\kf$ via
\eq{c2k}. We confine ourselves to the values of $|\kf|$ in the
range $0.1-10$, which we consider as natural - note that the
realization of FHI within quasi-canonical SUGRA requires
\cite{king,mur,hinova} $\kf$ significantly lower, i.e.,
$10^{-3}\lesssim\kf\lesssim0.01$. Then, for given $\ks$, we can
extract $c_{4K}$ through \eq{c4k}. In \Fref{fig1} we display the
resulting, this way, $\ck$ versus $\alpha_1$ for ${\sf M}=1$ and
$\ks=0$ (gray points) or ${\sf M}=2$ and $\ks=1$ (black points).
We present six families of points of different shapes
corresponding to different values of $\beta_1$ (gray points) or
$\alpha_2,\beta_1$ and $\beta_2$ (black points). The adopted
values for these parameters are shown in the r.h.s of \Fref{fig1}.
We observe that a wide range of negative $\ck$'s can be produced
with natural values of the parameters related to the structure of
K\"ahler potential ($\kf,~\ks,~\ai$ and $\bi$). As we verify below
(see \Sref{res}) these $\ck$'s assist us to achieve hilltop-type
FHI consistently with the data on $\ns$ for all possible values of
$\kappa$ or $M_{\rm S}$.

\section{The Inflationary Potential}
\label{sec:Vhi}

The general form of the potential which can drive the various
versions of FHI reads
\beq\label{Vol} V_{\rm HI}=V_{\rm HI0}\,\left(1+\ c_{\rm
HI}+\,{1\over4}\ck\sigma^4\right),\eeq
where, besides the contributions originating from $V_{\rm SUGRA}$
in \eq{Vfinals} (with $c_{2K}=0$), we include the term $c_{\rm
HI}V_{\rm HI0}$ which represents a correction to $V_{\rm HI}$
resulting from the SUSY breaking on the inflationary valley, in
the cases of standard \cite{susyhybrid} and shifted \cite{jean}
FHI, or from the structure of the classical potential in the case
of smooth \cite{pana1} FHI. $c_{\rm HI}$ can be written as
follows:

\begin{equation} \label{Vcor} c_{\rm HI}=\left\{\bem
{\kappa^2{\sf N}}\left[2 \ln\left(\kappa^2x M^2 /
Q^2\right)+f_c(x)\right]/32\pi^2\hfill
  & \mbox{for standard FHI}, \hfill \cr
{\kappa^2}\left[2 \ln\left(\kappa^2x_\xi M_\xi^2
/Q^2\right)+f_c(x_\xi)\right]/16\pi^2\hfill &\mbox{for shifted
FHI,} \hfill \cr
-2\mu_{\rm s}^2M_{\rm S}^2/27\sigma^4\hfill  &\mbox{for smooth
FHI}, \hfill \cr\eem
\right.\end{equation}
$$\mbox{with}\>\>
f_c(x)=(x+1)^{2}\ln(1+1/x)+(x-1)^{2}\ln(1-1/x)\Rightarrow
f_c(x)\simeq3\>\>\mbox{for}\>\>x\gg1,$$
$x=\sigma^2/2M^2$ and $x_\xi=\sigma^2/M^2_\xi$. Also ${\sf N}$ is
the dimensionality of the representations to which $\bar{\Phi}$
and $\Phi$ belong and $Q$ is a renormalization scale. Note that
renormalization group effects \cite{espinoza} remain negligible in
the available parameter space of our models.

For $\ck<0$, $V_{\rm HI}$ reaches a maximum at $\sigma=\sigma_{\rm
max}$ which can be estimated as follows:
\begin{equation}
\label{sigmamax} V'_{\rm HI}(\sigma_{\rm max})=0~~\Rightarrow~~
\sigma_{\rm max}\simeq\left\{\bem
\left(\kappa^2{\sf N}/8\pi^2\left|\ck\right|\right)^{1/4}\hfill &
\mbox{for standard FHI}, \hfill \cr
\left(\kappa^2/4\pi^2\left|\ck\right|\right)^{1/4}\hfill
&\mbox{for shifted FHI}, \hfill \cr
\left(8\mu_{\rm S}^2M_{\rm
S}^2/27\left|\ck\right|\right)^{1/8}\hfill &\mbox{for smooth FHI},
 \hfill \cr\eem
\right.\end{equation}
with $V''_{\rm HI}(\sigma_{\rm max})<0$, where the prime denotes
derivation w.r.t $\sigma$. The system can always undergo FHI
starting at $\sigma < \sigma_{\rm max}$. However, the lower
$n_{\rm s}$ we want to obtain, the closer we must set $\sigma_*$
to $\sigma_{\rm max}$, where $\sigma_{*}$ is the value of $\sigma$
when the scale $k_*$ crosses outside the horizon of FHI. To
quantify somehow the amount of this tuning in the initial
conditions, we define \cite{gpp} the quantity:
\beq \Dex=\left(\sigma_{\rm max} - \sigma_*\right)/\sigma_{\rm
max}.\label{dms}\eeq

\section{Observational Constraints}
\label{obs}

Under the assumption that (i) the curvature perturbations
generated by $\sigma$ is solely responsible for the observed
curvature perturbation and (ii) there is a conventional
cosmological evolution (see below) after FHI, the parameters of
the FHI models can be restricted imposing the following
requirements:

\begin{itemize}

\item The number of e-foldings $N_{\rm HI*}$ that the scale $k_*$
suffers during FHI is to account for the total number of
e-foldings $N_{\rm tot}$ required for solving the horizon and
flatness problems of SBB, i.e.,
\begin{equation}
\label{Nhi} N_{\rm HI*}=N_{\rm tot}~~\Rightarrow~~\:
\int_{\sigma_{\rm f}}^{\sigma_{*}}\, d\sigma\: \frac{V_{\rm
HI}}{V'_{\rm HI}}\simeq64.94+{2\over 3}\ln V^{1/4}_{\rm HI0}+
{1\over3}\ln {T_{\rm Hrh}}.
\end{equation}
where $\sigma_{\rm f}$ is the value of $\sigma$ at the end of FHI,
which can be found, in the slow-roll approximation, from the
condition
\beq \label{slow} {\sf max}\{\epsilon(\sigma_{\rm
f}),|\eta(\sigma_{\rm f})|\}=1,~~\mbox{where}~~
\epsilon\simeq{1\over2}\left(\frac{V'_{\rm HI}}{V_{\rm
HI}}\right)^2~~\mbox{and}~~\eta\simeq \frac{V''_{\rm HI}}{V_{\rm
HI}}\cdot \eeq
In the cases of standard \cite{susyhybrid} and shifted \cite{jean}
FHI, the end of FHI coincides with the onset of the GUT phase
transition, i.e., the slow-roll conditions are violated close to
the critical point $\sigma_{\rm c}=\sqrt{2}M$ [$\sigma_{\rm
c}=M_\xi$] for standard [shifted] FHI, where the waterfall regime
commences. On the contrary, the end of smooth \cite{pana1} FHI is
not abrupt since the inflationary path is stable w.r.t $\Phi-\bar
\Phi$ for all $\sigma$'s and $\sigma_{\rm f}$ is found from
Eq.~(\ref{slow}). On the other hand, the required $N_{\rm tot}$ at
$k_*=0.002/{\rm Mpc}$ can be easily derived \cite{hinova}
consistently with our assumption of a conventional
post-inflationary evolution. In particular, we assume that FHI is
followed successively by the following three epochs: (i) the
decaying-inflaton dominated era which lasts at a reheat
temperature $T_{\rm Hrh}$, (ii) a radiation dominated epoch, with
initial temperature $T_{\rm Hrh}$, which terminates at the
matter-radiation equality, (iii) the matter dominated era until
today.

\item The power spectrum $P_{\cal R}$ of the curvature
perturbations generated by $\sigma$ at the pivot scale $k_{*}$ is
to be confronted with the WMAP5 data~\cite{wmap}:
\begin{equation}  \label{Prob}
P^{1/2}_{\cal R}=\: \frac{1}{2\sqrt{3}\, \pi} \; \frac{V_{\rm
HI}^{3/2}\left(\sigma_*\right)}{|V'_{\rm
HI}\left(\sigma_*\right)|}\simeq\: 4.91\cdot
10^{-5}~~\mbox{at}~~k_*=0.002/{\rm Mpc}.
\end{equation}
\end{itemize}

Finally we can calculate the spectral index, $n_{\rm s}$, and its
running, $a_{\rm s}$, through the relations:
\begin{equation}\label{ns}
n_{\rm s}=\: 1-6\epsilon_*\ +\
2\eta_*\>\>\mbox{and}\>\>\alpha_{\rm s}
=\:{2\over3}\left(4\eta_*^2-(n_{\rm s}-1)^2\right)-2\xi_*,
\end{equation}
respectively, where $\xi\simeq V'_{\rm HI} V'''_{\rm HI}/V^2_{\rm
HI}$ and the variables with subscript $*$ are evaluated at
$\sigma=\sigma_{*}$.

We can obtain an approximate estimation of the expected $n_{\rm
s}$'s, if we calculate analytically the integral in
Eq.~(\ref{Nhi}) and solve the resulting equation w.r.t $\sigma_*$.
We pose $\sigma_{\rm f}=\sigma_{\rm c}$ for standard and shifted
FHI whereas we solve the equation $|\eta(\sigma_{\rm f})|=1$ for
smooth FHI ignoring any SUGRA correction. Taking into account that
$\epsilon<\eta$ we can extract $n_{\rm s}$ from Eq.~(\ref{ns}). We
find
\begin{equation} \label{nssugra} n_{\rm s}=\left\{\bem
1-{1/N_{\rm HI*}}+{3\kappa^2{\sf N}N_{\rm HI*}\ck/4\pi^2} \hfill &
\mbox{for standard FHI}, \hfill \cr
1-{1/N_{\rm HI*}}+{3\kappa^2N_{\rm HI*}\ck/2\pi^2} \hfill
&\mbox{for shifted FHI}, \hfill \cr
1-{5/3N_{\rm HI*}}+4\ck\left(6\mu^2_{\rm S}M^2_{\rm S}N_{\rm
HI*}\right)^{1/3}\hfill &\mbox{for smooth FHI}. \hfill \cr\eem
\right.\end{equation}
From the expressions above, we can easily infer that $\ck<0$ can
diminish significantly $n_{\rm s}$. To this end, in the cases of
standard and shifted FHI, $|\ck|$ has to be of order unity for
relatively large $\kappa$'s and much larger for lower $\kappa$'s
whereas, for smooth FHI, a rather low $|\ck|$ is enough.

\section{Numerical Results}
\label{res}

In our numerical investigation, we fix ${\sf N}=2$. This choice
corresponds to the left-right symmetric gauge group ${
SU(3)_c\times SU(2)_L\times SU(2)_R \times U(1)_{B-L}}$ for
standard FHI and to the Pati-Salam gauge group ${SU(4)_c\times
SU(2)_L\times SU(2)_R}$ for shifted \cite{jean} FHI. Note that, if
$\bar\Phi$ and $\Phi$ are chosen to belong to ${SU(2)_R}$ doublets
with $B-L=-1,~1$ respectively, no cosmic strings are produced
\cite{trotta} during this realization of standard FHI. As a
consequence, we are not obliged to impose extra restrictions on
the parameters (as, e.g., in Refs.~\cite{mairi, jp}). We also take
$T_{\rm Hrh}\simeq4\cdot10^{-10}$ (recall that the quantities with
mass dimensions are measured in units of $m_{\rm P}$) as in the
majority of these models \cite{lept, lectures, sstad} saturating
conservatively the gravitino constraint \cite{gravitino}. This
choice for $T_{\rm Hrh}$ do not affect crucially our results,
since $T_{\rm Hrh}$ appears in Eq.~(\ref{Nhi}) through the one
third of its logarithm and so its variation upon two or three
orders of magnitude has a minor influence on the value of $N_{\rm
tot}$.

The inflationary dynamics is controlled by the parameters (note
that we fix $c_{2K}=0$):
$$ \sigma_*, v_{_G}, \ck~~\mbox{and}\left\{\begin{matrix}
\kappa\hfill   & \mbox{for standard and shifted (with fixed
$M_{\rm S})$ FHI}, \hfill \cr
M_{\rm S}\hfill  &\mbox{for smooth FHI}. \hfill \cr\end{matrix}
\right.$$
In our computation, we can use as input parameters $\kappa$ or
$M_{\rm S}$, $\sigma_*$ and $\ck$. We then restrict $v_{_G}$ and
$\sigma_*$ so as Eqs.~(\ref{Nhi}) and (\ref{Prob}) are fulfilled.
Using Eq.~(\ref{ns}) we can extract $n_{\rm s}$ and $\alpha_{\rm
s}$ for any given $c_{4K}$ derived from \eqss{c2k}{c4k}{ab}.
Turning the argument around, we can find the observationally
favored $c_{4K}$'s, imposing the satisfaction of
Eq.~(\ref{nswmap}), and then we can check if these $c_{4K}$'s can
be derived from \eqss{c2k}{c4k}{ab}.

\begin{figure}[!t]\vspace*{-.21in}
\hspace*{-.25in}
\begin{minipage}{8in}
\epsfig{file=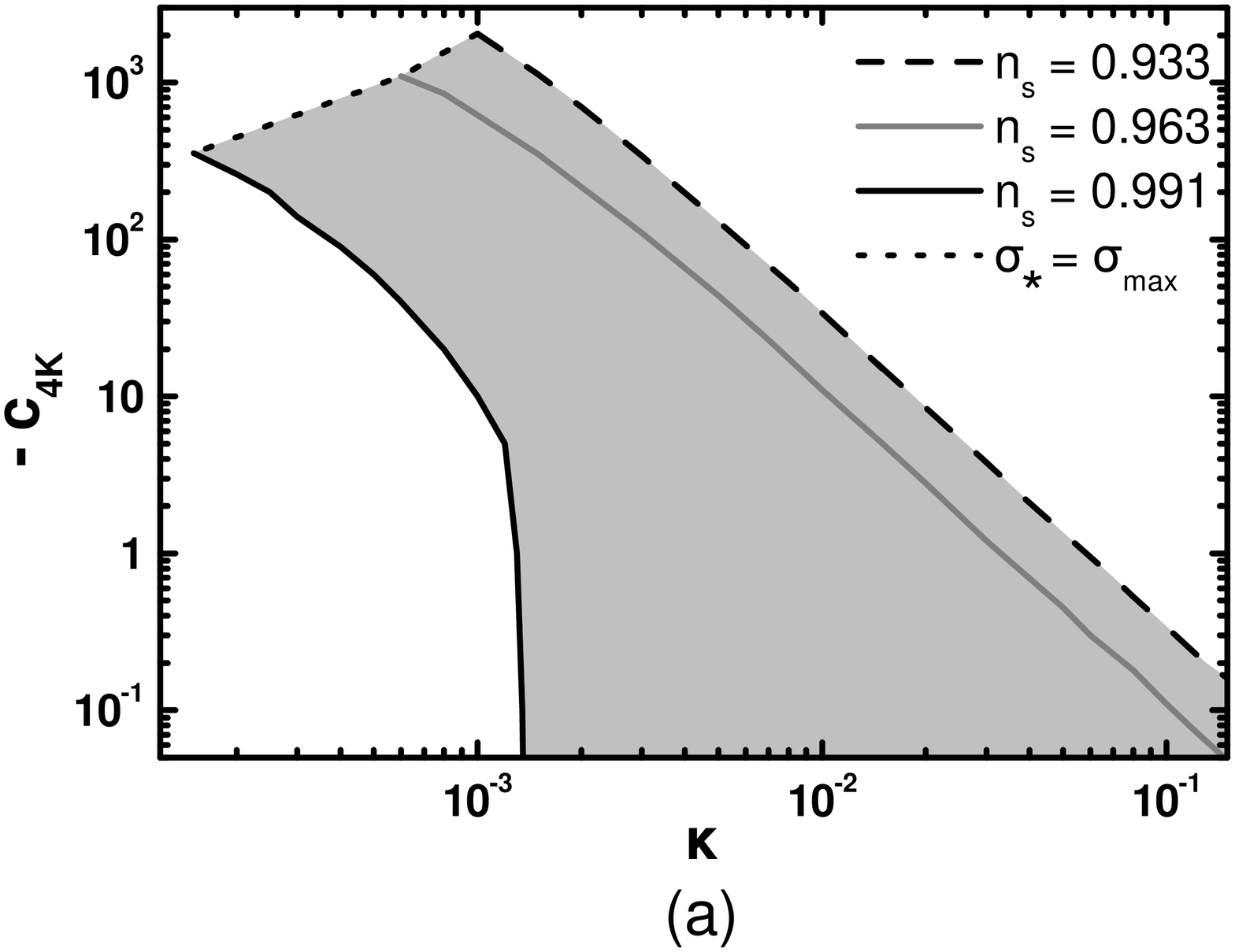,height=3.65in,angle=-90} \hspace*{-1.37 cm}
\epsfig{file=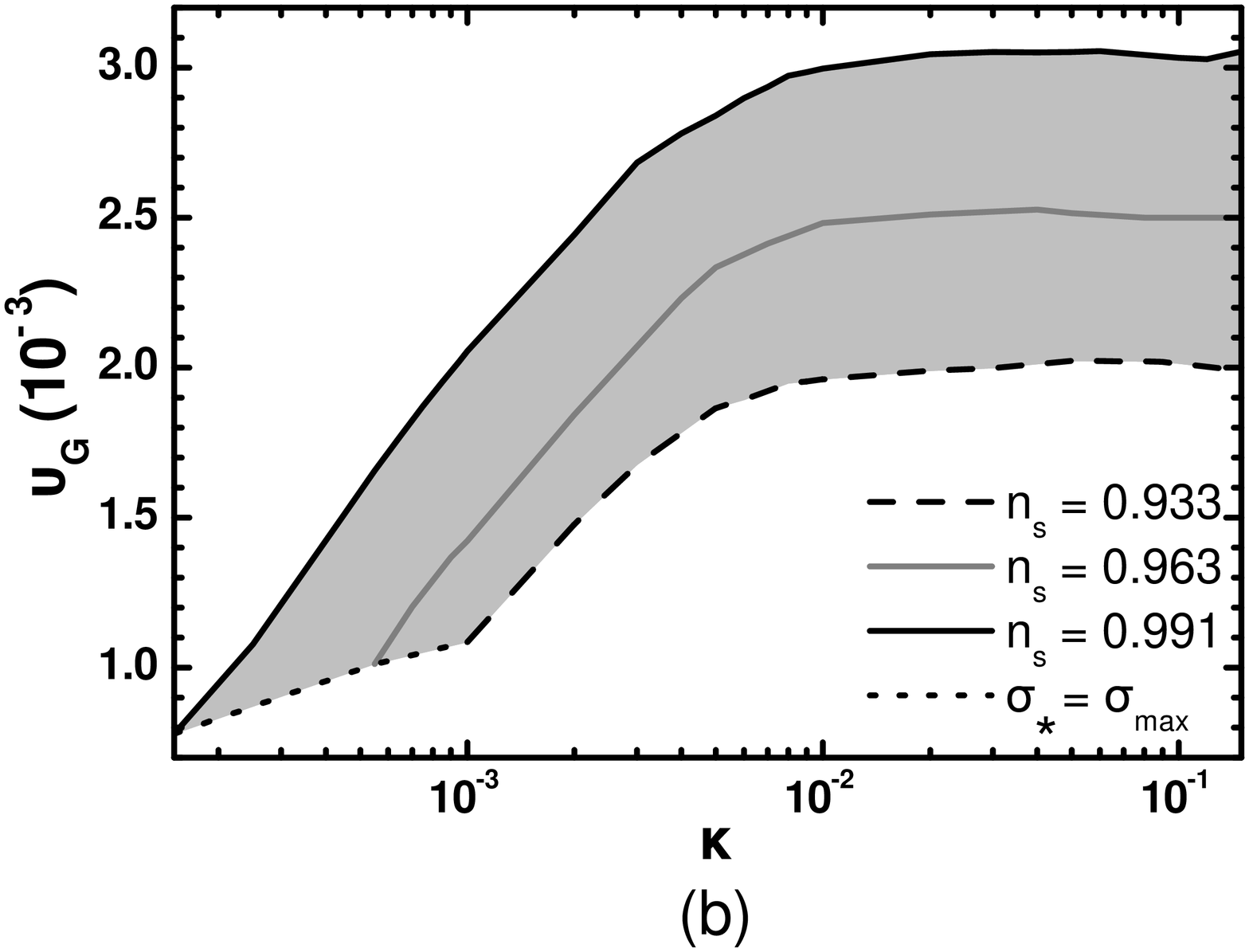,height=3.65in,angle=-90} \hfill
\end{minipage}
\hfill \vchcaption[]{\sl\small  Observationally allowed (lightly
gray shaded) areas in the $\kappa-c_{4K}$ [$\kappa-v_{_G}$] plane
(a) [(b)] for standard FHI with $c_{2K}=0$. The conventions
adopted for the various lines are also shown. }\label{fig2}
\end{figure}

Our results are presented in \Fref{fig2} for standard FHI and in
Table~\ref{table} for shifted and smooth FHI. Let us discuss these
results in the following.

\subsection{Standard FHI}  In \sFig{fig2}{a} [\sFig{fig2}{b}] we delineate the (lightly gray
shaded) regions allowed by Eqs.~(\ref{nswmap}), (\ref{Nhi}) and
(\ref{Prob}) in the $\kappa-\ck$ [$\kappa-v_{_G}$] plane for
standard FHI. The conventions adopted for the various lines are
also shown in the r.h.s of each graphs. In particular, the black
solid [dashed] lines correspond to $n_{\rm s}=0.991$ [$n_{\rm
s}=0.933$], whereas the gray solid lines have been obtained by
fixing $n_{\rm s}=0.963$ -- see Eq.~(\ref{nswmap}). Below the
dotted line, our initial assumption $\sigma_*<\sigma_{\rm max}$ is
violated. The various lines terminate at $\kappa=0.15$, since for
larger $\kappa$'s the two restrictions in \eqs{Nhi}{Prob} cannot
be simultaneously met. Note that for $n_{\rm s}=0.991$ and $
1.3\cdot10^{-3}\lesssim\kappa\lesssim0.15$ the curve is obtained
for positive $0\lesssim\ck\lesssim0.025$, not displayed in
\sFig{fig2}{a}.

From our data, we can deduce that (i) $v_{_G}$, $\ck$ and $\Dex$
increase with increasing $\ns$, for fixed $\kappa$ and (ii) $\ck$
and $\Dex$ increase with increasing $\kappa$, for fixed $\ns$. In
particular, for $n_{\rm s}=0.963$ we obtain
$$ 0.0006\lesssim\kappa\lesssim0.15\>\>\mbox{with}\>\>1.1\lesssim
v_{_G}/10^{-3}\lesssim2.5,\>-1100\lesssim
\ck\lesssim-0.05\>\>\mbox{and}\>\>0.014\lesssim\Dex\lesssim0.28.$$
Note that the $v_{_G}$'s encountered here are lower that those
found in the minimal SUGRA scenario (compare, e.g., with the
results of \cref{hinova}). Also, as in the case of quasi-canonical
SUGRA \cite{king,gpp}, a degree of tuning required for the values
of $\Dex$ in \eq{dms}. In particular for $\kappa>10^{-3}$, we find
$\Dex>10\%$. However, the situation becomes rather delicate as
$\kappa$ gets smaller than $10^{-3}$, for $\ns <0.97$. In this
case, $\Dex$ tends to zero, leading to a substantial tuning at the
few per cent level.

Comparing \Fref{fig1} and \sFig{fig2}{a}, we observe that the
required $\ck$'s, in order to achieve $\ns$'s within the range of
\eq{nswmap}, can be easily derived from the fundamental parameters
of the proposed K\"ahler potentials in \eqs{K}{k2}. Namely, for
$\ck<1$, ${\sf M}=1$ is sufficient, whereas $\ck>1$ necessitates
${\sf M}=2$ with $\beta_1$ and $\beta_2$ of different signs. It is
worth mentioning that even the rather large $\ck$'s can be
extracted from natural values of $\ai,~\bi,~\kf$ and $\ks$.

\subsection{Shifted and Smooth FHI}

In the cases of shifted and smooth FHI we confine ourselves to the
values of the parameters which give $v_{_G}=M_{\rm GUT}$ and
display solutions consistent with Eqs.~(\ref{Nhi}) and
(\ref{Prob}) in Table~\ref{table}. The selected $\ck$'s are
indicated in Table 1 (for $\ck\geq0$) and denoted by light gray
crosses in \Fref{fig1} (for $\ck<0$). The entries without a value
assigned for $\Dex$ mean that $V_{\rm HI}$ has no distinguishable
maximum.

We observe that the required (in order to obtain $v_{_G}=M_{\rm
GUT}$) $\kappa$'s in the case of shifted FHI are rather low and
so, reduction of $n_{\rm s}$ to the level dictated by
Eq.~(\ref{nswmap}) requires rather high $\ck$'s. These can be
derived, e.g., for ${\sf M}=2$ and $\bi$ of different signs. On
the contrary, in the case of smooth FHI, $n_{\rm s}$ turns out to
be quit close to its central value in \eq{nswmap} even with
$\ck=0$. Therefore, in order to reach the central and the lowest
value of $\ns$ in \eq{nswmap}, one needs rather small $\ck$'s,
which can be obtained even with ${\sf M}=1$ (and only negative
$\bi$'s) -- see \Fref{fig1}. However, the resulting $\Dex$'s are
lower than those of shifted FHI.

\setcounter{table}{1}
\renewcommand{\arraystretch}{1.2}
\begin{table}[!t]
\begin{center}
\begin{tabular}{|l|llll||l|llll|}
\hline
\multicolumn{5}{|c||}{\sc Shifted FHI}&\multicolumn{5}{|c|}{\sc
Smooth FHI}\\ \hline\hline
$\ck$ &${-50875\over1024}$&$-16$&$0$&${5}$&$\ck$&${-1127\over4096}$ & ${-1\over16}$&$0$&${3\over16}$\\
$\Dex/10^{-2}$ &  $13$&$26$&$-$&$-$&$\Dex/10^{-2}$ &
$5.5$&$17$&$-$&$-$
\\ \hline\hline
$\sigma_*/10^{-2}$ &$2.44$ &$2.29$&$2.2$&$2.17$&$\sigma_*/10^{-2}$ & $10$&$10.8$&$11$&$11.7$\\
$\kappa/10^{-3}$ & $8.33$&$8.8$&$9.23$&$9.4$&$M_{\rm S}/10^{-1}$ & $4.5$&$3.5$&$3.22$&$2.5$\\ \hline
$M/10^{-3}$&$9.15$& $9.31$&$9.44$&$9.5$&$\mu_{\rm S}/10^{-4}$& $3$&$3.9$&$4.3$&$5.45$\\
$1/\xi$ &  $4.19$&$4.28$&$4.36$&$4.4$&$\sigma_{\rm f}/10^{-2}$&\multicolumn{4}{|c|}{$5.5$}
\\
$N_{\rm HI*}$ &  $51.5$&$52.4$&$52.2$&$52.2$&$N_{\rm HI*}$ & $52.4$&$52.3$&$52.5$&$52.6$\\\hline
$n_{\rm s}$ &  $0.933$&$0.961$&$0.981$&$0.99$&$n_{\rm s}$ & $0.936$&$0.961$&$0.969$&$0.993$\\
$-\alpha_{\rm s}/10^{-4}$ &  $1.86$&$3.6$&$3.4$&$5$ &$-\alpha_{\rm
s}/10^{-4}$ & $4.5$&$5.3$&$5.8$&$7.7$\\ \hline
\end{tabular}
\end{center}
\vchcaption[]{\sl \small Input and output parameters consistent
with Eqs.~(\ref{Nhi}) and (\ref{Prob}) for shifted (with $M_{\rm
S}=0.205$) or smooth FHI, $v_{_G}=M_{\rm GUT}$ and selected
$\ck$'s indicated in Table~1 and \Fref{fig1}. }\label{table}
\end{table}

\section{Conclusions}\label{con}

We considered the basic types of FHI in the context of a string
inspired SUGRA scenario using a simple class of K\ähler potentials
given by \eq{K} with dependence -- see \eq{k2} -- on the hidden
sector fields. We imposed, essentially, two conditions so that
hilltop FHI can be realized. Namely, we required the mass squared
of the inflaton during FHI is zero and the parameter $\ck$
involved in the quartic SUGRA correction to the inflationary
potential is adequately negative so that the results on $n_{\rm
s}$ can be reconciled with data. We found a wide and natural set
of solutions which satisfy the above requirements. Moreover the
desired form of the K\ähler potential is thus obtained for all
hidden sector v.e.vs and not just for some carefully chosen
vaccua. However, our results require a proximity between the
values of the inflaton field at the maximum of the potential and
at the horizon crossing of the pivot scale. The amount of this
tuning was measured by the quantity $\Dex$ defined in \eq{dms}. In
particular, for $n_{\rm s}$ close to its central value, we found
that (i) in the case of standard FHI, $v_{_G}$ turns out to be
well below $M_{\rm GUT}$ with $\ck\simeq-(1100-0.05)$ for
$\kappa\simeq(0.0006-0.15)$ and $\Dex\simeq(1.4-28)\%$; (ii) in
the case of shifted [smooth] FHI, we succeeded to obtain
$v_{_G}=M_{\rm GUT}$ for $\ck\simeq-16$ [$\ck\simeq-1/16$] and
$\Dex=26\%$ [$\Dex=17\%$]. Observationally less interesting
$\ns$'s can be also achieved for $\ck\geq0$, without the presence
of a maximum along the inflationary trajectory.

Trying to compare our construction with that of \cref{nurmi} we
would like to mention that in our case (i) there is no need for
cancellation of the term $V_0$ in the expansion of \eq{Vfinal};
(ii) higher order terms of the inflaton in the K\"ahler potential
let intact our calculation since only terms up to the order $\p^4$
in the inflationary potential are relevant for our analysis; (iii)
the requirement of the $h_m$'s stabilization before the onset of
FHI can be evaded if $\W=0$. In the latter case, $h_m$ can
represent even fields of the observable sector which do not
contribute to the superpotential at all, due, e.g., to the
existence of an additional symmetry (as in the case of
\cref{sugraP}).

Throughout our investigation we concentrated on the predictions
derived from the inflationary potential, assuming that we had
suitable initial conditions for FHI to take place. In general, it
is not clear \cite{king, mur} how the inflaton can reach the
maximum of its potential in the context of hilltop inflation.
Probably an era of eternal inflation prior to FHI could be useful
\cite{lofti} in order the proper initial conditions to be set. On
the other hand, in our regime with $\ck<0$, the potential develops
just a maximum along the inflationary path and not a local maximum
and minimum as in the case with quasi-canonical K\"ahler potential
\cite{king,gpp,mur,hinova}. Therefore, in our scheme,
complications related to the trapping of the inflaton in that
local minimum are avoided.

Let us finally note that a complete inflationary model should
specify the transition to radiation domination, and also explain
the origin of the observed baryon asymmetry. For FHI with
canonical or quasi-canonical K\"ahler potential, this has been
extensively studied (see, e.g., Ref. \cite{jean, pana1, lept, gpp,
king, sstad}). Obviously our set-up preserves many of these
successful features of this post-inflationary evolution which may
constrain further the parameter space of our models and help us to
distinguish which version of FHI is the most compelling.

\begin{acknowledgement}
The author would like to thank cordially G. Lazarides for helpful
discussions and S.~Nurmi for useful correspondence. This research
was funded by the FP6 Marie Curie Excellence Grant
MEXT-CT-2004-014297.
\end{acknowledgement}

\end{document}